# The Roman Colonia Marciana Ulpia Traiana Thamugadi (Timgad) and Trajan’s Birthday

**Amelia Carolina Sparavigna**
Politecnico di Torino

It is told that the Roman Colonia Marciana Ulpia Traiana Thamugadi, that is Timgad in Algeria, had been oriented to the sunrise on the day of Trajan’s birthday, that is 18 September 100 AD. Here we use software CalSKY to investigate the sunrise azimuth and compare it to the direction of the decumanus of the Roman town.

In 2012, I wrote an article concerning the orientation of Trajan's town of Timgad [1]. Timgad was a Roman colonial town founded by the Emperor Trajan in AD 100. The Roman full name was “Colonia Marciana Ulpia Traiana Thamugadi”: in this name we find the names of emperor's mother Marcia, his father Marcus Ulpius Traianus and his eldest sister Ulpia Marciana. The ruins of the Trajan's Timgad are in Algeria. It the book entitled "Ancient Town-Planning", written by F. Haverfield and published in 1913 [2], Timgad is proposed as a noteworthy site for being one of the best examples of the Roman city planning. When the author wrote the book, about Trajan's Timgad there were only purely archaeological remains. Haverfield reports that the ruins are on "the northern skirts of Mount Aurès, halfway between Constantine and Biskra and about a hundred miles from the Mediterranean coast. Here the emperor Trajan founded in A.D. 100 a 'colonia' on ground then wholly uninhabited, and peopled it with time-expired soldiers from the Third Legion which garrisoned the neighbouring fortress of Lambaesis ... The 'colonia' of Trajan appears to have been some 29 or 30 acres in extent within the walls and almost square in outline (360 x 390 yds.). It was entered by four principal gates, three of which can still be traced clearly, and which stood in the middle of their respective sides; the position of the south gate is doubtful. According to Dr. Barthel, the street [decumanus] which joins the east and west gates was laid out to point to the sunrise of September 18, the birthday of Trajan." [2,3].
In [1], I used an approach based on a simple equation, to discuss the affirmation given in [2], about the direction of the decumanus. Here I would like to repeat the same, using software CalSKY (this software had been used in [4]). First, let us consider the site of Timgad and apply software on equinoxes (Table 1, Pag.3) and winter solstice (Table 2, Pag.4). We do not show the results for the summer solstice, because the same sunrise azimuth is spread on several days. In the Tables 1 and 2, we can see data given by CalSKY for 100 AD and 2019 AD. Since the dates of spring equinoxes and winter solstices have a difference of one day, we are sure that we have not to worry about the calendar. Then, let us consider the day of 18 September of 2019 and 100 AD. In the Table 3 at Pag.5 we can see the results given by software. Please consider that the tables are giving the astrological sign, that is, the sector of the twelve 30° sectors of the ecliptic - from the vernal equinox -, in which we find the sun.
Let us consider the sunrise azimuths and the altitude of the sun in the following Table 4.

| | | | |
|---|---|---|---|
| **18 Sep 2019** | Rise : 6h18.7m az= 87.0° | Set : 18h37.4m az=272.8° | Transit: 12h28m21s Altitude=56.4° Vir |
| | Dawn : 5h23m | Dusk : 19h33m | Day : 12h18.8m |
| **18 Sep 100** | Rise : 5h29.9m az= 86.4° | Set : 17h51.3m az=273.3° | Transit: 11h40m51s Altitude=56.9° Vir |
| | Dawn : 4h34m | Dusk : 18h47m | Day : 12h21.4m |

Table 4. Courtesy CalSKY. Sunrise azimuth (astronomical horizon) on 18 September 2019 and 100 AD.

The sunrise azimuth (without considering the local natural horizon and the effect of the atmospheric refraction) was of 86.4° on 100 A, and it is of 87.0° on 2019. So, a difference of half a degree exists. The same we see for the altitude of the sun. Let us consider the azimuth 86.4° and compare to the direction of the decumanus. Here we use satellite images. Of course a local measurement is necessary for further detailed studies. From a Bing Map we have the following Screenshot 1.

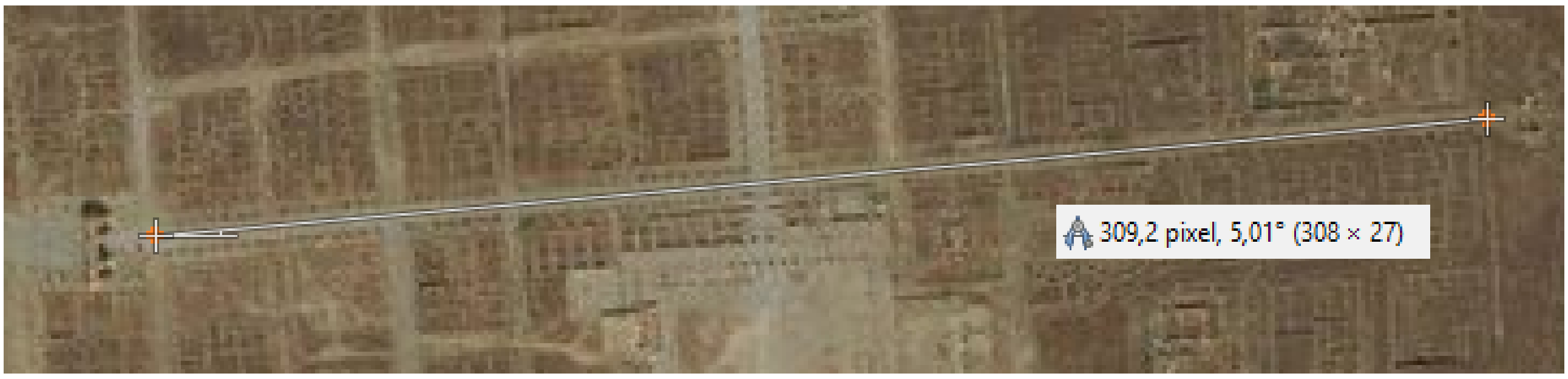

Screenshot 1: Satellite image courtesy Bing Maps. The azimuth of the decumanus is of 85°.

In the screenshot you can see the measurement of the angle made by means of GIMP pair of compasses. The azimuth of the decumanus, from true North, is of 85°. As a conclusion, using CalSKY, we have that it is possible that Timgad had been founded on 18 September 100 AD. On that day the sunrise azimuth was of about 86.5°. We have a difference of about 1.5°. However, it is possible to reduce this difference, including the effect of atmospheric refraction. Let us assume that the azimuth could be affected of 0.5°, then the sunrise azimuth can be estimated as 86°, and the difference reduced to a degree. This is a result which is far better than that given in [1]. Of course, the result here proposed is given for an astronomical horizon. Using Google Earth, we can see a simulated horizon without relevant hills in the direction of the decumanus.

I would like to conclude in the following manner. Assuming the site of Timgad laying on a flat surface, which is the same of that observed in Screenshot 1, we could use data given above to estimate the uncertainty of the directions of the decumani, planned by means of a centuriation method. It could be of about one degree. Actually we do not know if the surveyor used the first ray of the sun he perceived or the full sun to determine the direction of the decumanus. It is meaningless to give conclusions concerning the days of the foundation, based on smaller uncertainties of the direction of decumani. Of course GPS investigations at the archaeological sites can given us astonishing results, but the precision of the modern measurements cannot be transferred to the ancient methods, without a careful evaluation. In any case, it seems to me that it is impossible to determine the specific year of the foundation by comparing the azimuths of the decumani to the sunrise azimuths. And this is clearly shown by the case of Timgad. To change the difference of half a degree we need two thousand years. Therefore, from a sunrise azimuth we cannot determine the year of foundation.

## References

[1] Sparavigna, A. C. (2012). The orientation of Trajan's town of Timgad. arXiv preprint arXiv:1208.0454.

[2] Haverfield, F. (1913). Ancient town - planning, Oxford, The Clarendon Press, 1913, available at http://www.gutenberg.org/files/14189/14189-h/14189-h.htm

[3] Barthel, W. (1911). Römische Limitation in der Provinz Africa, 1911, CXX, pp. 39-126. Carl Georgi Verlag, Bonn

[4] Sparavigna, A. C. (2016). Astronomy and Tidal Analysis Applied to the Study of Julius Caesar's Commentarii (August 2, 2016). Available at SSRN: https://ssrn.com/abstract=2817327 or http://dx.doi.org/10.2139/ssrn.2817327

| | | | |
|---|---|---|---|
| **20 Mar 100** | Rise : 5h52.6m<br>az= 90.2° | Set : 17h57.5m<br>az=270.0° | Transit: 11h54m45s<br>Altitude=54.0° Psc |
| | Dawn : 4h58m | Dusk : 18h53m | Day : 12h04.9m |
| **21 Mar 100** | Rise : 5h51.1m<br>az= 89.7° | Set : 17h58.2m<br>az=270.5° | Transit: 11h54m23s<br>Altitude=54.4° Psc |
| | Dawn : 4h56m | Dusk : 18h53m | Day : 12h07.1m |
| **19 Mar 2019** | Rise : 6h39.9m<br>az= 90.2° | Set : 18h44.7m<br>az=270.0° | Transit: 12h42m00s<br>Altitude=54.0° Psc |
| | Dawn : 5h45m | Dusk : 19h40m | Day : 12h04.8m |
| **20 Mar 2019** | Rise : 6h38.4m<br>az= 89.7° | Set : 18h45.5m<br>az=270.5° | Transit: 12h41m42s<br>Altitude=54.4° Psc |
| | Dawn : 5h43m | Dusk : 19h41m | Day : 12h07.1m |

Table 1: Courtesy CalSKY. Spring equinox 2019 AD and 100 AD at Timgad.

| | |
|---|---|
| **20 Dec 2019** | Rise : 7h38.7m az=118.5° Set : 17h24.3m az=241.4° Transit: 12h31m32s Altitude=31.1° Sgr |
| | Dawn : 6h38m Dusk : 18h25m Day : 9h45.6m |
| **21 Dec 2019** | Rise : 7h39.2m az=118.6° Set : 17h24.8m az=241.4° Transit: 12h32m02s Altitude=31.1° Sgr |
| | Dawn : 6h39m Dusk : 18h25m Day : 9h45.6m |
| **22 Dec 2019** | Rise : 7h39.7m az=118.6° Set : 17h25.3m az=241.4° Transit: 12h32m31s Altitude=31.1° Sgr |
| | Dawn : 6h39m Dusk : 18h26m Day : 9h45.6m |
| **23 Dec 2019** | Rise : 7h40.2m az=118.6° Set : 17h25.8m az=241.4° Transit: 12h33m01s Altitude=31.1° Sgr |
| | Dawn : 6h40m Dusk : 18h26m Day : 9h45.6m |
| **24 Dec 2019** | Rise : 7h40.7m az=118.5° Set : 17h26.4m az=241.5° Transit: 12h33m31s Altitude=31.1° Sgr |
| | Dawn : 6h40m Dusk : 18h27m Day : 9h45.7m |
| **25 Dec 2019** | Rise : 7h41.1m az=118.5° Set : 17h27.0m az=241.5° Transit: 12h34m01s Altitude=31.1° Sgr |
| | Dawn : 6h41m Dusk : 18h27m Day : 9h45.8m |
| **20 Dec 100** | Rise : 6h56.5m az=118.9° Set : 16h40.4m az=241.1° Transit: 11h48m29s Altitude=30.9° Sgr |
| | Dawn : 5h56m Dusk : 17h41m Day : 9h43.9m |
| **21 Dec 100** | Rise : 6h57.1m az=118.9° Set : 16h40.9m az=241.1° Transit: 11h49m00s Altitude=30.9° Sgr |
| | Dawn : 5h57m Dusk : 17h41m Day : 9h43.8m |
| **22 Dec 100** | Rise : 6h57.6m az=118.9° Set : 16h41.4m az=241.1° Transit: 11h49m30s Altitude=30.9° Sgr |
| | Dawn : 5h57m Dusk : 17h42m Day : 9h43.8m |
| **23 Dec 100** | Rise : 6h58.1m az=118.9° Set : 16h42.0m az=241.1° Transit: 11h50m01s Altitude=30.9° Sgr |
| | Dawn : 5h58m Dusk : 17h42m Day : 9h43.9m |
| **24 Dec 100** | Rise : 6h58.5m az=118.9° Set : 16h42.5m az=241.2° Transit: 11h50m31s Altitude=30.9° Sgr |
| | Dawn : 5h58m Dusk : 17h43m Day : 9h44.0m |

Table 2: Courtesy CalSky. Winter solstice at Timgad in 2019 AD and 100 AD.

| | | | |
|---|---|---|---|
| **16 Sep 2019** | Rise : 6h17.2m<br>az= 86.0° | Set : 18h40.4m<br>az=273.7° | Transit: 12h29m04s<br>Altitude=57.2° Leo |
| | Dawn : 5h22m | Dusk : 19h36m | Day : 12h23.2m |
| **17 Sep 2019** | Rise : 6h17.9m<br>az= 86.5° | Set : 18h38.9m<br>az=273.3° | Transit: 12h28m43s<br>Altitude=56.8° Leo |
| | Dawn : 5h23m | Dusk : 19h34m | Day : 12h21.0m |
| **18 Sep 2019** | Rise : 6h18.7m<br>az= 87.0° | Set : 18h37.4m<br>az=272.8° | Transit: 12h28m21s<br>Altitude=56.4° Vir |
| | Dawn : 5h23m | Dusk : 19h33m | Day : 12h18.8m |
| **19 Sep 2019** | Rise : 6h19.4m<br>az= 87.4° | Set : 18h36.0m<br>az=272.3° | Transit: 12h28m00s<br>Altitude=56.0° Vir |
| | Dawn : 5h24m | Dusk : 19h31m | Day : 12h16.5m |
| **20 Sep 2019** | Rise : 6h20.2m<br>az= 87.9° | Set : 18h34.5m<br>az=271.8° | Transit: 12h27m38s<br>Altitude=55.6° Vir |
| | Dawn : 5h25m | Dusk : 19h30m | Day : 12h14.3m |
| **16 Sep 100** | Rise : 5h28.2m<br>az= 85.4° | Set : 17h54.1m<br>az=274.3° | Transit: 11h41m27s<br>Altitude=57.7° Vir |
| | Dawn : 4h33m | Dusk : 18h50m | Day : 12h26.0m |
| **17 Sep 100** | Rise : 5h29.0m<br>az= 85.9° | Set : 17h52.7m<br>az=273.8° | Transit: 11h41m09s<br>Altitude=57.3° Vir |
| | Dawn : 4h34m | Dusk : 18h48m | Day : 12h23.7m |
| **18 Sep 100** | Rise : 5h29.9m<br>az= 86.4° | Set : 17h51.3m<br>az=273.3° | Transit: 11h40m51s<br>Altitude=56.9° Vir |
| | Dawn : 4h34m | Dusk : 18h47m | Day : 12h21.4m |
| **19 Sep 100** | Rise : 5h30.7m<br>az= 86.9° | Set : 17h49.8m<br>az=272.9° | Transit: 11h40m34s<br>Altitude=56.5° Vir |
| | Dawn : 4h35m | Dusk : 18h45m | Day : 12h19.1m |
| **20 Sep 100** | Rise : 5h31.5m<br>az= 87.4° | Set : 17h48.4m<br>az=272.4° | Transit: 11h40m16s<br>Altitude=56.1° Vir |
| | Dawn : 4h36m | Dusk : 18h44m | Day : 12h16.9m |

Table 3: Courtesy CalSKY. September at Timgad in 2019 AD and 100 AD.